\documentclass[prl,showpacs,floatfix,amsfonts]{revtex4}
\usepackage{graphicx,graphics,color,epsfig}
\usepackage{bm}
\usepackage{amsmath}
\usepackage{amssymb}
\newcommand{\bS}{{\bf S}}

\newcommand{\bk}{{\bf k}}

\newcommand{\bsig}{\mbox{\boldmath{$\sigma$}}}
\newcommand{\beqa}{\begin{eqnarray}}
\newcommand{\eeqa}{\end{eqnarray}}
\newcommand{\bI}{\bf{I}_s}

\newcommand{\wL}{\omega_{L}}

\begin{document}
\preprint{}
\title{1/f Spin Noise and a Single Spin Detection  with STM}

\author{Yishay Manassen$^1$ and  A. V. Balatsky$^2$} \affiliation{$^1$
 Department of Physics and the Ilse Katz
Center for Nanometer Scale Science and Technology, Ben Gurion
University, Beer Sheva, 84105, Israel; $^2$ Theoretical Division,
Los Alamos National Laboratory, Los Alamos, New Mexico 87545 USA }

\date{April 22, 2003}

\begin{abstract}
{ We propose a novel mechanism for single spin detection based on
the 1/f spin current noise. We postulate the 1/f spin noise for
the tunneling  current, similar to the ubiquitous  1/f noise in
magnetic systems. Magnetic coupling between tunneling electrons
and localized spin ${\bf S}$ then leads to the peak at Larmor
frequency in the power spectrum of the electric current
fluctuations $I^2_{\omega}$. The elevated noise in the current
spectrum will be spatially localized near the magnetic site. The
difference in the power spectra taken at the Larmor frequency and
elsewhere would reveal the peak in the spectrum. We argue that
signal to noise ratio for this mechanism is on the order one.

In addition we discuss the asymmetric lineshapes observed
regularly with this measurement. We show that such lineshapes are
in accordance to the random sampling done with the tunneling
electrons. Yet, this predicts a linewidth at least one order of
magnitude larger than observed experimentally which is likely to
be due to electrostatic repulsion between the tunneling electrons
and temporal correlations in the tunneling process. }
\end{abstract}
\pacs{ 76.30.-v, 07.79.Cz, 75.75.+a}

\maketitle

\section{1/f spin noise}

   The phenomenon of 1/f (flicker) noise is known for almost 80 years
\cite{1}. It describes  the deviation from the flat spectral
density expected from a current made of uncorrelated charge
carriers - at low frequencies. In this range, the spectral density
was found to obey a power law of the form $1 /f^\alpha$ where $f$
is the frequency, and $\alpha = 0.5 - 1.5 $. Flicker noise appears
in endless number of electronic devices, in music \cite{2} in
ocean streams \cite{3,4} and in many entirely different systems.
This is one of the most universal phenomena, yet, one of the
largest enigmas in physical sciences.

   An early explanation to this phenomenon was that the 1/f noise
can arise from a superposition of relaxation processes \cite{5}.
In this model the noise is described as a superposition of
consecutive random events, each starts at a certain time $t_0$ and
follow a simple exponential relaxation law: $N(t-t_0)=N_0 e^{-
(t-t_0)/\tau}$. The power spectrum of one such event is a
Lorentzian. The power spectrum of a large number of such
consecutive random events, all with the same $\tau$ is also a
Lorentzian. If, on the other hand, there is a distribution of
relaxation times $P(\tau) \sim 1/\tau$, from $\tau_1$ to $\tau_2$
than the overall spectral density will obey a power law $\sim 1/f$
in the range between $\tau^{-1}_1$ and $\tau^{-1}_2$. A common
denominator to many mechanisms proposed for different 1/f
phenomena \cite{6,7} is a distribution of relaxation times.

 $1/f$ noise is ubiquitous in STM tunneling current although its
origins are a source of continuous mystery that is not fully
understood until now \cite{8,9,10,11}.  Unlike the well known shot
noise $1/f$ noise is proportional
    to the square of the current such that: $ \langle I^2(\omega)\rangle / I^2 = const$.
    Where $\langle I^2(\omega) \rangle$  is the spectral density of
    current fluctuations (in units of A$^2$/Hz) and I is the current. Empirically,
     the current fluctuations in 1/f noise are known to obey the Hooge formula
     \cite{12} $\langle I^2(\omega) \rangle /I^2 =  a 1/[fN]$ where N are the number
     of current carriers  in the sample and $a$
     is of the order of $ 0.01$. The appearance of $1/f$ noise in the STM is a
      surprising observation because the normal "explanation" of this noise
       is a fluctuating defect with a wide distribution of relaxation times.
        Such explanations are not suitable here, because of the extremely local
         nature of the measurement. The STM measurements of $1/f$ noise
(in a voltage of $-0.5V$) gives a peak at zero frequency where the
intensity of the current noise $\langle I^2(\omega) \rangle$ is 20
times larger than the Thermal (Nyquist) noise which will give the
order of magnitude of the noise in higher frequencies (at room
temperature) and amounts to $41fA/\sqrt{Hz} $ . The width of the
observed $1/f$ noise peak at zero frequency is of the order of
10-100KHz.

   A special case of 1/f noise relevant to the present paper
is the magnetic flicker noise. In this case the noise is due to
fluctuations in the magnetization. Normally, it is assumed that
these fluctuations constitute an additional noise source in any
magnetic system. However, low frequency magnetic noise is hard to
observe experimentally and sensitive detection techniques and
special systems are required. Such fluctuations were detected for
the first time in spin glasses with a SQUID magnetometer
\cite{13,14}. They  were detected with SQUID also in
antiferromagnetic thin films \cite{15} and on superparamagnetic
nanoparticles \cite{16}.
   Other ways to detect 1/f magnetic noise is by following the
resistance fluctuations close to a certain ferromagnetic
transition (colosal magnetoresistance) \cite{17}, and to observe
the electrical noise generated in small Hall probe contacts
\cite{18}. The 1/f noise generated in the Hall contacts is much
larger than the usual 1/f noise measured in the same system which
is unrelated to magnetic fluctuations. Despite the failure to
measure spatial correlations it is obvious \cite{18} that there is
a certain coherence length such that two noise measurements done
at a distance smaller than this length will give the same results.

Although we have no direct experimental evidence we claim that
since conduction electrons in metal constitute also a (special)
magnetic system, there should be 1/f magnetic fluctuations also in
the spins in such a system which is also paramagnetic. ESR of
conduction electrons is known for many years \cite{19,20}. These
spectra are known for their special lineshape \cite{21} and for
the narrow lines ($T_1$ and $T_2$ are equal and large). We would
like to emphasize that the sequence of individual dephasing events
which are responsible for $T_2$ relaxation of conduction electrons
are very similar to the sequence of relaxation events which are
responsible for 1/f noise in general. Therefore, there are
distributions of relaxation times that are  expected to give 1/f
magnetic fluctuations in conduction electrons in metals.

Such 1/f magnetic noise is expected to be of general significance
in spintronics applications. We discuss here the implications of
such 1/f magnetic noise (either in the spins of the conduction
electrons or in a regular paramagnetic system) on single
precessing spin detection with STM (ESR-STM). It will be shown
below that this 1/f spin noise explains the observation of the
signal from non magnetic tips and elucidates several aspects in
this technique. We find  that the interaction  of the tunneling
electrons spin  with the local impurity spins  simply couples the
1/f magnetic noise  with the  noise of the local spin $\bf{S}$
thus creating a peak in the current noise at  Larmor frequency.

  ESR-STM is a technique that is using the extremely local
   nature of the STM measurement to detect the precession of
 isolated spin centers on the surface. When a tip of an STM
 is located above a paramagnetic spin center (in the presence
 of an external magnetic field) the tunneling current is
  modulated by the precession. It was shown \cite{22,23} that the
  AC current at the Larmor frequency is spatially localized
   within $0.5-1nm$. It is the spatial localization that indicates
   (though it must be proved) that this technique is capable
    of detecting a single spin. In addition it was proved that
     the frequency of the signal is dependent on real time on
     the size of the magnetic field \cite{24,25}. More recently
similar experiments have been done on the paramagnetic BDPA
molecule \cite{26}. A recent paper shows that ESR-STM can be done
also on a TEMPO molecule, revealing the hyperfine spectrum \cite
{39}. The interest in this technique has risen sharply recently,
due to the possibility to manipulate and detect a single spin
\cite{27,33} and due to the possibility to use it for quantum
computation \cite{27,28}.
   There have been many proposals for the mechanism of this phenomenon
\cite{29,30,31,32,33,34,35,36}.
   However, an experimental verification for any of the proposals is still required.

    In our previous papers \cite{30,31} we discussed the
     following question: What is the role of the Heisenberg
      exchange interaction in ESR-STM?  Under which circumstances can
      a tip emitting tunneling electrons with a random spin orientation create an
       elevated noise level at the Larmor freuqency through
        interaction with the single precessing spin?

It was argued in \cite{30,31} that existence of long time
correlations in the temporal spin polarization of tunneling
current is sufficient to provide the elevated noise at the Larmor
frequency in the current noise. Here we further build upon this
idea and show  that the $1/f$ noise in the tunneling spin  current
is sufficient to produce the effect. We stress that no
correlations in the spin polarization of the tunneling electrons
within the precession period is required to produce the elevated
noise at Larmor frequency in this approach.

We discussed the spin dependent tunneling matrix element:
 \beqa
 G=
G_0\exp[- [(F-J\bS(t)\cdot{\bf s}(t))/F_0]]. \eeqa
  $s$ is a spin matrix with implicit spin indices that are
  given by conduction electron spin operator $ s^i = 1/2 \sigma^i_{\alpha,\beta}, i =
x,y,z$ , $\bS(t)$ is the local impurity spin. In an external
magnetic field it will have a random dynamics with remnant of the
precession at the Larmor frequency. $F$ is the barrier height
(typically ~4eV), $J$ is the exchange coupling and $F_0$ is the
energy related to the distance $d$ between the tip and the surface
$ F_0=d^2/8md^2$. In this term, the tunneling probability depends
on the relative orientation between $\bS$ and ${\bf s}$.  We
write:
 \beqa G= G_0\exp[- [(F/F_0]] (cosh[(JS/2F_0)+{\bf s}(t)\cdot{\bf n}(t)sinh(JS/2F_0).
\eeqa
 \label{eq:G}
Where ${\bf n}$ is a unit vector in the direction of $\bS$.

 The Hamiltonian we consider describes a spin
dependent tunnneling matrix element between the tip (L electrode)
and the surface (R electrode)
\begin{equation}\label{eq:Ham}
   H = \sum_{\bk,\alpha} \epsilon(\bk) c^{\dag}_{L
\alpha}(\bk) c_{L
  \alpha}(\bk) + (L \rightarrow R) + \sum_{\bk,\bk'}
c^{\dag}_{L \alpha}(\bk)G_{\alpha \beta}
  c_{R \beta}(\bk')
\end{equation}
 We assume  that the magnetic field is  along z axis:
$B||z$ with corresponding Larmor frequency $\omega_L = g \mu_B B$.
The tunneling current operator will contain the spin independent
part that we omit hereafter and the spin dependent part:
\begin{equation}\label{eq:J1}
 \delta \hat{I}(t) = G_1 {\bf n}(t) {\bf I}_s(t),
\end{equation}
where $ G_1 = G_0\exp[- [(F/F_0]] \sinh[\frac{JS}{2F_0}]$, and
${\bf I}_s(t)$ is a spin current between tip and substrate. The
  dc current at a given bias $V$ is : $I_0 =g_0 V$, $g_0 = G_0\exp[- [(F/F_0]]
cosh[JS/2F_0]$. The current-current
 correlator, normalized to dc current  is then:
\begin{eqnarray}\label{eq:J2}
  \frac{\overline{\langle \delta \hat{I}(t) \delta
\hat{I}(t')\rangle}}{I^2_0} = (\sinh[\frac{JS}{2F_0}])^2
\nonumber \\
\sum_{i,j = x,y,z}
  \langle n^i(t)n^j(t')\rangle \overline{I_s^i(t)
I_s^j(t')}
\end{eqnarray}

More rigorous treatment, similar to the one done in \cite{37},
where one takes into account the effect of the local spin on the
mass current in the ignored terms will lead to the contribution of
the same order as the term we are focusing on.  Hence we use a
simplified formula for the current that gives the right order of
magnitude estimate.

The change in the tunneling conductance due to exchange
interaction between the tunneling electrons and the localized spin
leads to Eq(4): $
 \delta I(t) \sim  \bI(t)\cdot \bS(t)$.
 Only the transverse component contribute to this term ($S_x(t)$ and $S_y(t)$). When the
spins of the tunneling electrons are completely uncorrelated we
can say that the dispersion of the tunneling current which is
dependent on the spin (over one precession period T) is: $\sum_{i}
\langle \delta I^2\rangle \sim \sum_{i=1}^N
[s_x(t_i)I_{s,x}(t_i)]^2 + (x \rightarrow y) \sim <N>. $ Or, the
relative dispersion compared with the tunneling current will be:
$<I^2>/I^2  \sim 1/<N>$. To estimate the magnitude of the spin
dependent dispersion, namely the size of the noise which is due to
the interaction with the precessing spin, we use:
 \beqa ( <I^2 >)^{1/2}
/I_0 \sim 2/N sinh{[(JS/2F_0)}. \eeqa
 If we take a typical value
of $J=0.1eV, F=4eV, F_0=0.1eV $(for $d=4\AA$) and $S=1/2$, we get
a value of $0.01$, namely rf intensity of 10 picoampere, which is
within the right order of magnitude.

\begin{figure}
\epsfig{figure=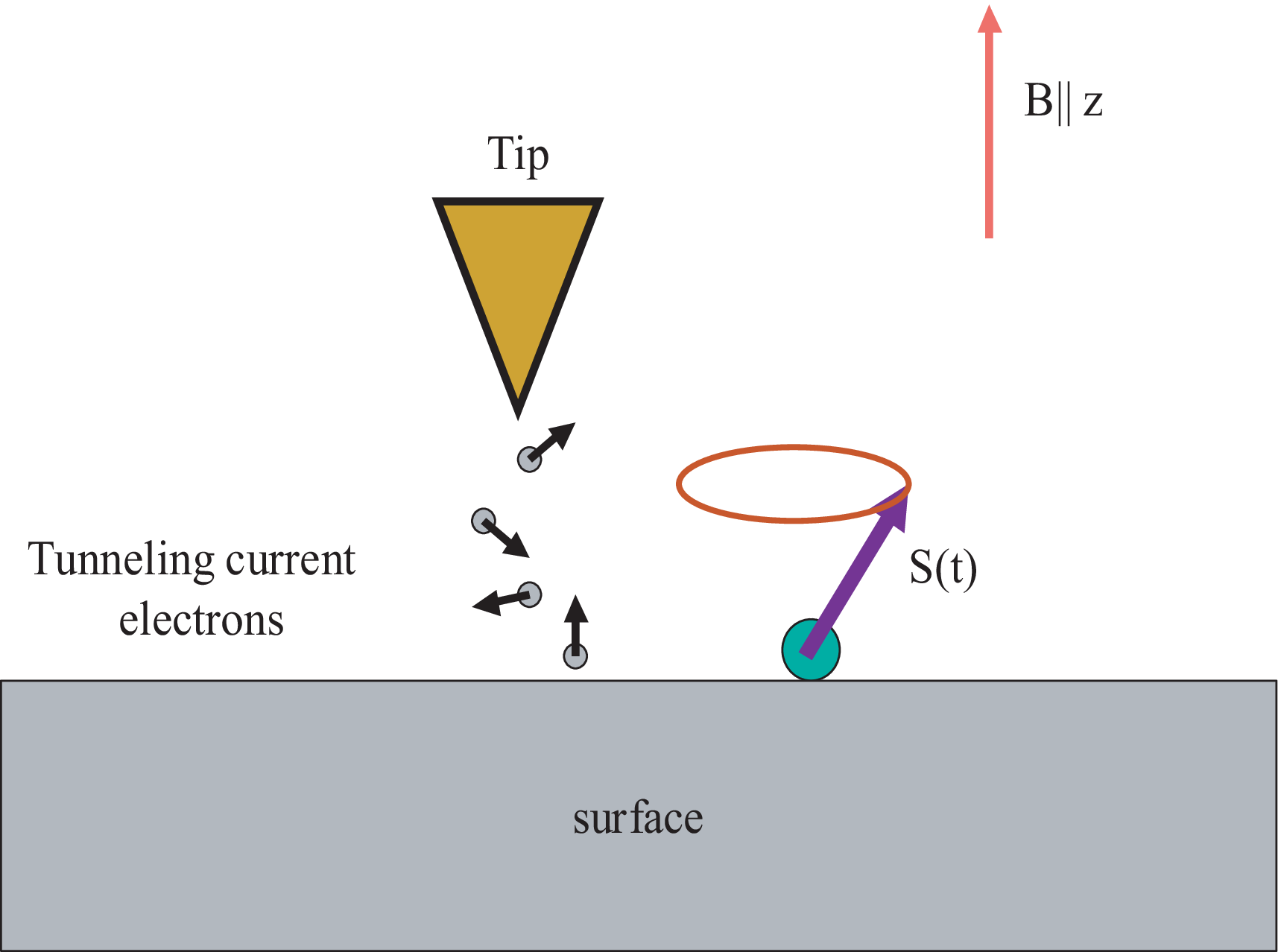,width=3in}
\caption[*]{  Schematics of the ESR-STM experiment is shown. The
fluctuations in the spin polarization of the tunneling electrons
at the time scale of the precession $T$ will be nonzero and will
scale as $\frac{1}{\overline{N}}$.   Once the tip is positioned
close to the localized spin, the exchange interaction between the
localized spin and the tunneling electrons will modulate the
tunneling current. The conditions in which this random modulation
of the spin tunneling current with $1/f$ spectrum will create a
$\omega_L$ peak are discussed in the text.} \label{FIG:STM1}
\end{figure}

The spectrum is detected at the frequency domain, and by looking
at the Fourier transform of the current fluctuations. This means
that the spectrum is a convolution of the power spectrum of the
single spin fluctuations:
 $<S^2_i(\omega)> = \frac{\gamma}{[(\omega- \wL)^2 + \gamma^2]}$
and the power spectrum of the spin tunneling current: $
<I_{s,i}(\omega)I_{s,j}(-\omega)> = \delta_{ij}<I_s^2(\omega)>$.
Here, to be specific, we assumed the spectrum of a localized spin
to be Lorenzian, however the argument is applicable to the case of
general spectrum $<S^2_i(\omega)>$.
\begin{eqnarray}\label{eq:Ipower1}
 \frac{\overline{\langle I^2_{\omega} \rangle}}{I^2_0} =
  (\sinh[\frac{JS}{2F_0}])^2
\nonumber \\
\sum_{i = x,y,z} \int \frac{d\omega_1}{2\pi} \langle
(I_s^i)^2_{\omega - \omega_1}\rangle \langle
(S^i)^2_{\omega_1}\rangle
\end{eqnarray}
If the tunneling electron  spins are completely uncorrelated, then
the power spectrum is white noise, and the expected signals will
be scattered over the whole frequency range. In order to get a
peak at the Larmor frequency, some correlation at zero frequency
is required, namely the tunneling electron spins have to have a
some temporal  spin polarization in the long time limit \cite{35}.

   To give an example  we consider the case where several magnetic atoms are
   adsorbed on the tip. Then the spins of the tunneling electrons
   will be under the influence of slowly changing spins ${\bf h}_i$ of
   magnetic atoms due to exchange interaction $H_{int} = \sum_i
   J_i
   {\bf h}_i(t) \bsig (t) =  {\bf H}(t) \bsig(t)$ where $J_i$ are the exchange couplings
to the i-th
   atom and  we define ${\bf H}$ as a sum of random  fluctuating magnetic
   moments  ${\bf
   h}_i$
of individual magnetic impurities adsorbed on the tip. Here we
assume that only the "field" value ${\bf H}$ at
   the end of the tip matters as the tunneling electrons are
   affected by this field the most before tunneling out of the tip. This fluctuating
exchange field will
    spin polarize the tunneling current.  The correlator is
   $<{\bf H}(t){\bf H}(t')> = \Sigma_i <{\bf h}_i(t) {\bf h}_i(t')> $. The correlation
function from
   each impurity assuming they are independent, obeys $<{\bf h}_i (t) {\bf h}_i (t')> =
e^{-(t-t')/ \tau_i }$.
   Average over distribution times $\tau_i$ will with
   probability distribution $P(\tau) \sim 1/\tau$ between $\tau_1$ and
   $\tau_2$, will yield the 1/f noise for the field fluctuation power spectrum. This
    follows directly from evaluating the Lorenzian in frequency space $\frac{\tau}{\omega^2 \tau^2 + 1}$
    , averaged over distribution of relaxation time $\tau$ with $P(\tau)$; e.g see Ref[7]:
   \beqa
\langle{\bf H^2}(\omega)\rangle = \int_{\tau_1}^{\tau_2} d\tau
\exp (i \omega (t-t')) P(\tau)
    \langle {\bf h}_i (t) {\bf h}_i (t') \rangle \sim \frac{1}{\omega}
\eeqa
 Hence the tunneling current will acquire spin polarization fluctuations.
    It is natural to assume that the spin fluctuations of the tunneling electrons will be
proportional to fluctuating
    exchange field ${\bf H}$. The degree of spin polarization of the
    tunneling electron is  proportional to the Zeeman energy $(g \mu_B {\bf H})/W$
related to the total
     bandwidth $W \sim 1/N_0$.
    We get
    \beqa
{\bf I}_s(t)  =  I  g \mu_B {\bf H}(t) N_0
 \eeqa
 here we assumed
that the effect is proportional to the tunneling current $I$.
Indeed tunneling electrons will sample the random field ${\bf H}$
at a
 rate at which they tunnel and this rate is given by the electric current
 $I$.

  We now consider the specific case when spin  tunneling
    current $\bI$ has   $1/f$ noise component.  \beqa
 \langle
I^2_s(\omega) \rangle = \frac{C}{|\omega|}, C = a I^2
\label{EQ:1/fcurrent}\eeqa
 for all components
$i= x,y,z$. The magnitude of the $1/f$ noise is to be given by
$I^2$ up to an unknown numerical factor $a$ \cite{12}, where $N$
is absorbed in $a$ now.   $1/f$ noise is  peaked at zero frequency
and will provide a peak  at the Larmor frequency in the convolved
spectrum, Eq.(\ref{eq:Ipower1}). We also remark here that in a
similar fashion one can get the  $1/f$ noise in the spin current
due to $1/f$ noise in the
 electronic current in a presence of a constant polarizing field.
 We will not consider this possibility here.

  Spin tunneling
 current  will  have white noise asymptotic at   high frequencies. At
 lower frequencies, similar to the unpolarized electric current \cite{8,9,10,11},
 the relaxation processes that control spin relaxation in the
 current will contribute. To give a physically plausible argument
 about the origin of the $1/f$ noise consider  relaxation time of the tunneling spin
current:
  \beqa
\langle I_{s,i}(t)I_{s,j}(t')\rangle =
\delta_{ij}\exp(-|t-t'|/\tau) \eeqa this correlation function
leads to the Lorenzian for the noise spectrum with the width
$\gamma = 1/\tau$: $\langle I^2_{s, x}(\omega)\rangle =
\frac{\gamma}{\pi(\omega^2 + \gamma^2)}$.  Now we assume that
there is a distribution of relaxation times with probability
distribution $P(\tau) d\tau$. The possible origins of the
distribution of relaxation times in tunneling current can be the
multiple other spins in the vicinity and in the tip that produce
spin relaxation with different times. To obtain $1/f$ noise in the
current correlator one has to assume that $P(t)\sim 1/\tau$ in
some window of relaxation times, as was mentioned earlier.

The power spectrum of the electric current, using
Eq.(\ref{EQ:1/fcurrent})
 is:
\begin{eqnarray}\label{eq:Ipower3}
  & \frac{\langle I^2_{\omega} \rangle}{I^2} =
  (\sinh[\frac{JS}{2F_0}])^2
 \sum_{i = x,y,z} \int \frac{d\omega_1}{2\pi} \frac{1}{|\omega -
\omega_1|} \langle(S^i)^2_{\omega_1}\rangle \simeq \nonumber\\ &
(\sinh[\frac{JS}{2F_0}])^2 \frac{1}{[(\omega - \omega_L)^2 +
\gamma^2]^{1/2}}
\end{eqnarray}
Where we used the low frequency asymptotics for spin current Eq.
(\ref{EQ:1/fcurrent})  and assumed Lorenzian for spin correlation
function. We get finally :
 \beqa \label{eq:result} \frac{\langle
I^2_{\omega} \rangle}{I^2} \simeq a/\pi \frac{1}{((\delta
\omega)^2 + \gamma)^{1/2}} \eeqa Where we introduced the detuning
parameter $\delta \omega = \omega - \omega_L$.

The signal to noise ratio is controlled by the parameter $a$ and
is:
 \beqa
\frac{\langle I^2_{\omega} \rangle}{I^2} \simeq \frac{a}{\gamma
\pi}
 \eeqa
  Without the microscopic
model it is impossible to know exactly what is the magnitude of
$a, N, \epsilon$.  Self consistent treatment would give  the
scattering rate $\gamma \sim 10^6 Hz$ that  will be determined by
the maximum of the intrinsic decay due to backaction or the
extrinsic decay due to environment. As a guidance we  take
available STM data for the tunneling current noise
\cite{8,9,10,11}. In these data the $1/f$ noise is clearly seen
above the high frequency noise floor below 100kHz. From Hodge
formula we estimate $a \sim 10^{-2}$. We take $N \sim 10$ similar
to mass current case, as estimated in \cite{11}. It is convenient
to relate the power spectrum of the current fluctuation at the
peak at $\omega = \omega_L$ to the shot noise power spectrum
$\langle I^2_{shot}(\omega) \rangle = e I$. S/N ratio is in this
case is bound from above by the number on the order of unity

 \beqa
  \frac{\langle I^2_{\omega} \rangle}{\langle
I_{shot}^2(\omega)\rangle} \simeq  \frac{a}{N \pi \gamma \tau_e}
\sim O(1) \eeqa where we wrote explicitly the dependence of the
current fluctuation on $N$.

\section{Random Sampling}

In the previous section we discussed the conditions in which the
random orientation of the spins of the tunneling electrons can,
through interaction with a single precessing spin, give a signal
at the Larmor frequency. Nevertheless, this is only part of the
picture. Since the tunneling electrons are probing a periodic
precessional motion adiabatically, sequential tunneling and
temporal correlations are essential, as will be discussed below,
for getting a signal with a narrow linewidth.

One of the most important characteristic of the signals detected
with ESR-STM is their linewidth. the linewidth might reflect both
the lifetime of the spin state, and the back action effect of the
tunneling electrons on the precessing spin. However, it is
necessary to discuss an additional source of the linewidth which
is directly related to the way in which we probe the precession.
We claim here that the periodic precession of the spin is sampled
by the tunneling electrons. For each electron the precessing spin
looks static (Adiabatic process). The simplest case which will be
discussed here, is that the tunneling times of the tunneling
electrons are uncorrelated. (As will be shown this is not a
realistic assumption). In this case we can say that the sampling
times obey an exponential distribution (Poisson process). We show
here using the 1/f spin noise model described above that this is
in accordance with the asymmetric lineshapes observed in the
experiments. Also such a sampling process, should lead to a rapid
increase in the linewidth when the magnetic field is increased.
The data published so far, show such a trend.

We recall that in a tunneling current of 1nA there are
$6.25\times10^9$ electrons/sec. The measurements were performed in
a frequency of $2-8 \times 10^8$. This implies a ratio ($R_t$) of
$0.033-0.13$  of the average time between two tunneling electrons
and the precession time. According to the Nyquist sampling theorem
it is impossible to perform proper sampling of a periodic function
if the sampling dwell time is above half of the time that it takes
to complete one period. This is a fundamental limitation because
it makes no sense to increase the field such that the precession
frequency rises above $3\times 10^9$ Hz (for a current of one
nanoampere). However it is anticipated that this should be
important also in much smaller frequencies. Due to the
uncorrelated nature of the tunneling current, it is expected that
for a periodic function that is sampled in random, (according to
the Poisson distribution) increasing the frequency means that more
and more sampling times (between two consecutive tunneling events)
will be larger than half of the period. This implies that at
larger fields (or smaller currents) we expect that the linewidth
will increase. Moreover even in one peak we anticipate that the
high frequency side will be broader than the low frequency side,
which implies an asymmetric lineshape that has a larger slope on
the low frequency side. This is precisely the behavior that we see
in the measurements.

\begin{figure}
\epsfig{figure=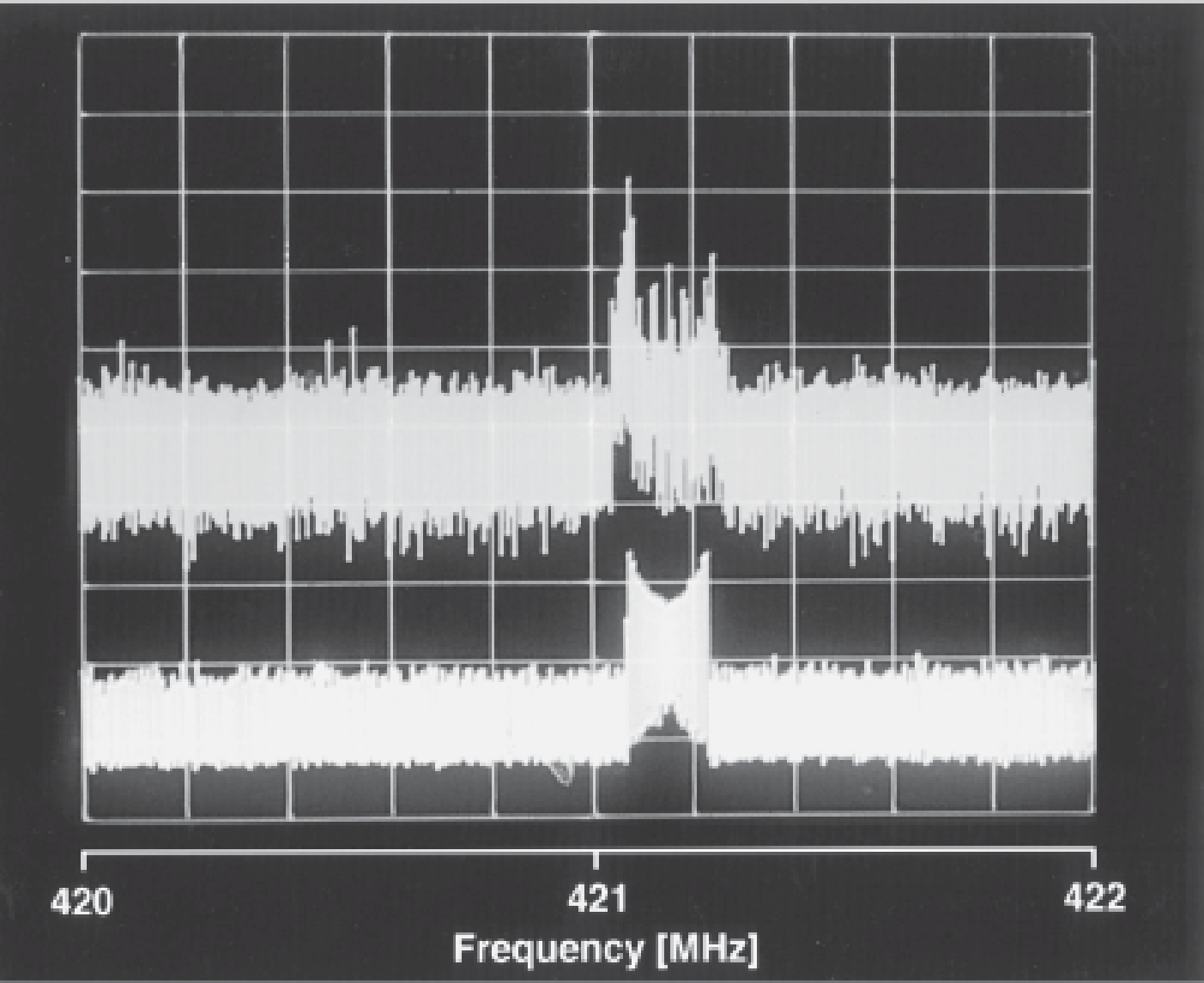,width=3in} \caption[*] {A frequency
modulated signal in a field of 150G. The field modulation
parameters are $\Delta$ H = 27 mG and the modulation frequency is
300KHz. Two spectra are presented. The upper figure represents the
data  from the STM, lower is simulation. It is clear that the
original line-shape is not symmetric. Taken from \cite{24}
}{\label{FIG:STM4}}
\end{figure}

\begin{figure}
\epsfig{figure=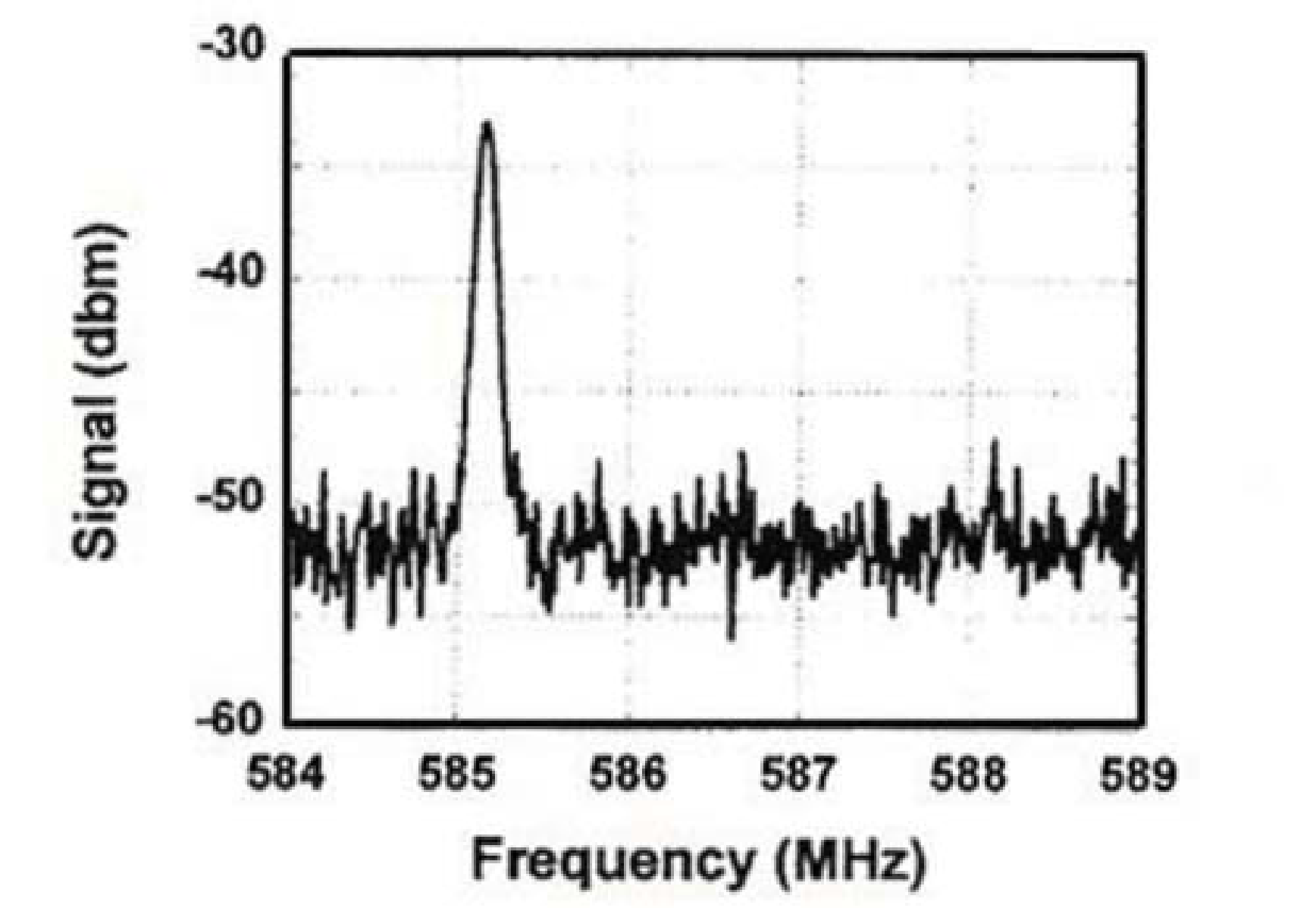,width=3in} \caption[*]{An asymmetric
ESR-STM spectrum of a BDPA cluster in a magnetic field of 210 G.
Taken from \cite{26}} \label{FIG:STM5}
\end{figure}

    Figs. 2,3 shows two spectra that were published already. Fig. 2
shows a frequency modulated signal that is taken from Ref.
\cite{24}. The splitting in this signal was used to prove that the
frequency depends on real time on the value of the magnetic field.
However it is easy to see that the original spectrum is asymmetric
and distorted. A similar distortion is seen in the (already
published) spectrum (Fig. 3). This spectrum was taken from Ref.
\cite{26}. The shape of the spectrum is completely asymmetric and
looks very similar to the Poisson distribution function. As
discussed in details in Ref. \cite{25}, this asymmetric lineshape
is the cause of getting an absorption lineshape when phase
sensitive detection is applied. The asymmetry was explained there
as a rapid passage phenomenon. As will be shown here, we believe
now that it can be explained by the physical nature of the
measurement. It is important to emphasize that this is not always
the case. Namely other lineshapes were also observed due to
spectral diffusion. However, this represents the  majority of
cases.

In order to explain this trend we have done very simple
simulation. In this simulation we have estimated the value
represented by Eq. 8. We have taken a sinusoidal function for $S$
and a random function for $I_s$ which has a long time correlations
(1/f peak).  However, in this simulation this function is sampled
in times that are determined by the exponential distribution.
After that we estimated the power spectral density in the observed
signal using the Welch method of spectral estimation \cite{38}.

\begin{figure}
\epsfig{figure=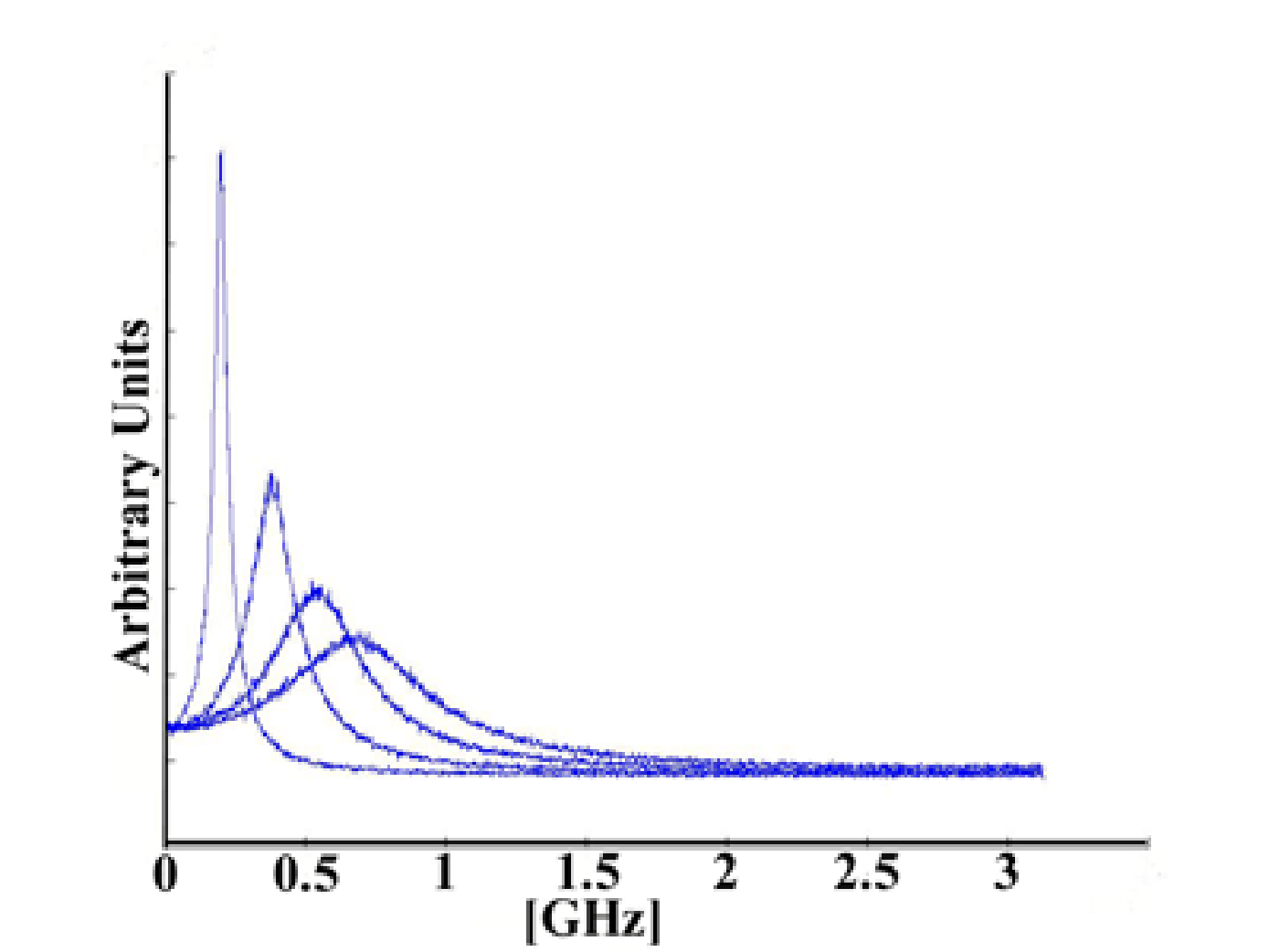,width=3in} \caption[*]
 {The simulated
lineshapes for a periodic function that is accumulated with a
Poisson distribution of sampling times. A current of 1nA was
assumed. The lineshape is clearly asymmetric. \label{FIG:STM2}}
\end{figure}

\begin{figure}
\epsfig{figure=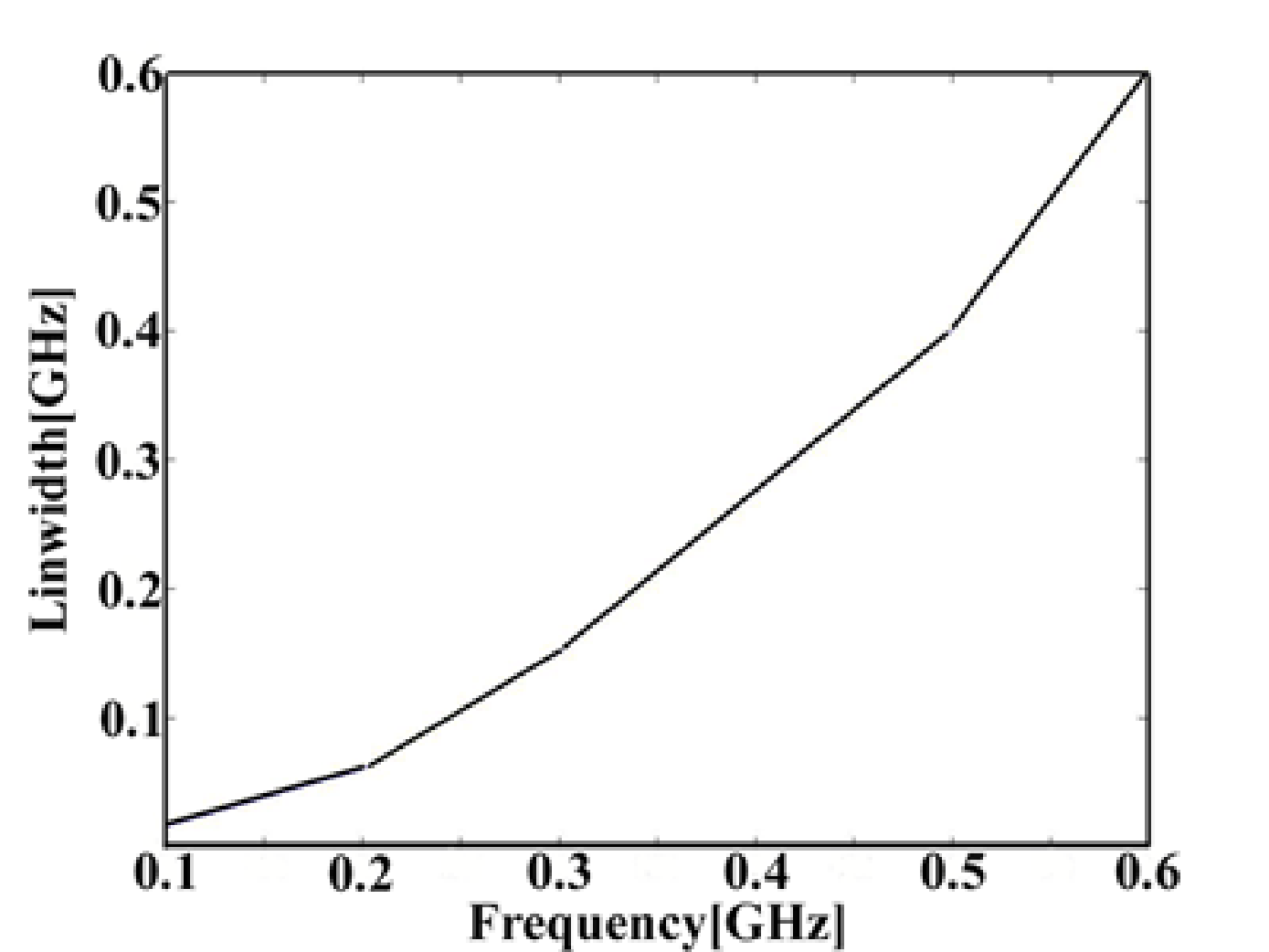,width=3in} \caption[*]{The predicted
linewidths as a function of the magnetic field. \label{FIG:STM3}}
\end{figure}

This was done in a current of 1nA and at the frequencies relevant
for the experiment. Each spectrum was taken for total sampling
time of 0.001 seconds. By changing the frequency we changed the
average number of electrons for each period. For each such
spectrum the power spectrum was calculated. Fig. 4 shows the
result of such a simulation for cases relevant to the experimental
situation. It is clear that the calculated linewidths are much
larger than anticipated from lifetime considerations. The
asymmetric lineshapes observed are very similar to those observed
from the experiment (compare with Fig. 3). A rapid increase in the
linewidth is predicted (Fig. 5). The data we have so far support
this prediction but more experiments are required (Fig.6). Using
longer lifetimes gives narrower lineshapes but the overall
behavior is the same. We want to emphasize that we do not take
into account the backaction effect of the tunneling electrons on
the precessing spin. A similar behavior is expected when the
tunneling current is reduced. However this may affect other
things, and we think that the dependence of the linewidth on the
field is a more informative measurement.

The lineshape and linewidth which are shown in Figs. 4 and 5 were
calculated with the 1/f spin noise model, but the same results
will be observed from other models as well. Sampling of a
sinusoidal function at times given by an exponential distribution
will give similar results.

It is clear that the calculated linewidths are at least one order
of magnitude larger than the experimental linewidths. This
difference reveals that the exponential distribution of sampling
times is totally unrealistic. Because of the small size of the
tunneling region in the STM tip, two electrons can not tunnel at
the same time because of the strong electrostatic repulsion
between them. Thus the exponential distribution has to be modified
such that some temporal correlations will be introduced in the
tunneling times. We are currently trying to get more data on the
linewidths in different magnetic fields, currents and bias
voltages, with the hope that ESR-STM lineshapes and linewidths
will reveal details on the temporal correlations in the tunneling
process.

\begin{figure}
\epsfig{figure=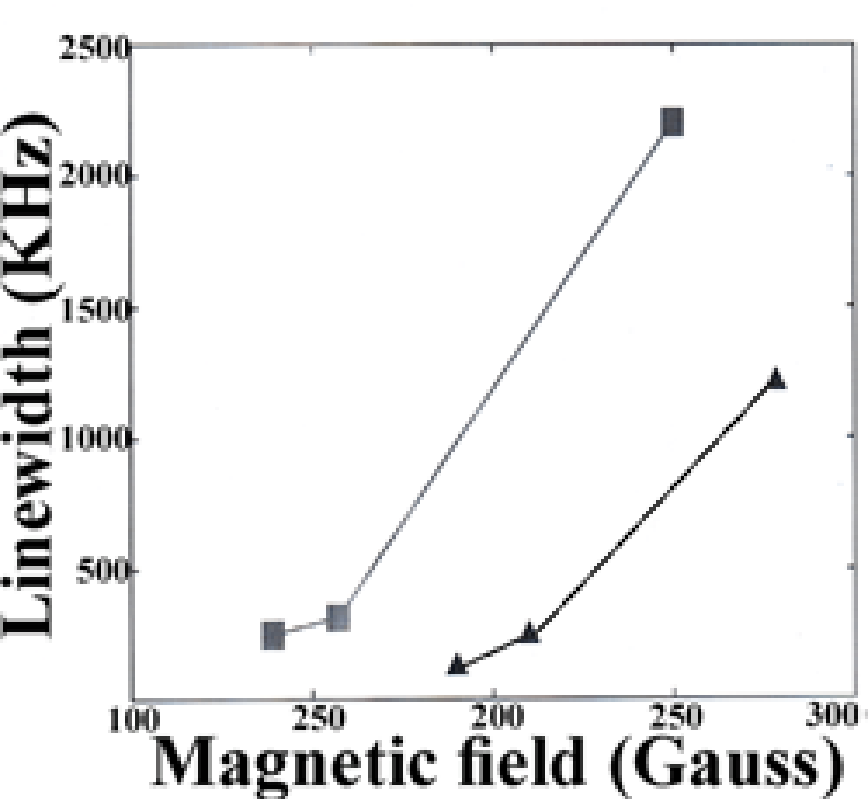,width=3in} \caption[*]{The linewidths as
a function of the magnetic field. The upper curve is the
linewidths measured for Si $P_b$ center \cite{23} while the lower
curve for BDPA molecules \cite{39} \label{FIG:STM4}}
\end{figure}

 {\bf Conclusion}

 We have proposed a 1/f noise as a  mechanism
of the coupling of  tunneling current from the STM tip to
localized spin ${\bf S}$. This mechanism allows one to detect a
signal from single spin even in the case when there is {\em no dc
spin polarization of the tunneling current}. Instead we argue that
the electric current will have a contribution coming from the
coupling of spin fluctuating current to the local spin. The best
way to detect a single spin in this approach is to perform a
difference experiment where the noise spectum is taken at the
local spin site and then at the non-magnetic site elsewhere on the
surface. The difference of two noise spectra would reveal the
localized spin contribution. We find the signal to noise ratio for
this model to be on the order S/N $\sim 1$.
   We discussed the affect of random sampling times on the
linewidths. The results explain the assymmetric lineshapes
commonly observed with ESR-STM. The results predict a rapid
increase of the linewidth with the field. The linewidths
calculated are too broad which indicates that there must be
correlations between the times of the tunneling events due to
Coulomb repulsion. We think that linewidth measurements can
provide a lot of information on the temporal nature of the
tunneling process.

{\bf Acknowledgement}

We are grateful to Z. Nussinov, J.X. Zhu, A. Shnirman for useful
discussions. This work was supported by a grant from the German
Israel Binational Science Foundation and LDRD at Los Alamos.


{\bf References}


\begin{thebibliography}{}




\bibitem{1} J. B. Johnson, $Phys.$ $Rev.$ {\bf 26}, 71 (1925).

\bibitem{2}   R. F. Voss and J. Clarke, $Nature$ {\bf 258}
317 (1975).

\bibitem{3}   B. A. Taft $et. al.$ $Deep$  $Sea$  $Research$ {\bf
21} 403 (1974).

\bibitem{4}   C. Wunsch, $Rev.$  $Geophys.$ $and$ $Space$ $Phys.$
{\bf 10} 1, (1972).

\bibitem{5}   W. Schottky, $Phys.$ $Rev.$ {\bf 28} 74,
(1926).

\bibitem{6}   P. Dutta, P. Dimon and P. M. Horn, $Phys.$
$Rev.$ $Lett.$ {\bf 43} 646 (1979).

\bibitem{7}  P. Dutta and P. M. Horn, $Rev.$ $Mod.$ $Phys.$ {\bf 53}, 497
(1981).

\bibitem{8}   R. M\"{o}ller, A. Esslinger and B. Koslowski, $Appl.$
$Phys.$ $Lett.$ {\bf 55}, 2360 (1989).

\bibitem{9}   R. M\"{o}ller, A. Esslinger, and B. Koslowski, $J.$ $Vac.$
$Sci.$ $Technol.$ $A$ {\bf 8}, 590 (1990).

\bibitem{10} R. M\"{o}ller, C. Baur, A. Esslinger and P. K?rz, $J.$ $Vac.$
$Sci.$ $Technol.$ $B$ {\bf 9}, 609 (1991).

\bibitem{11}   K. Maeda, S. Sugita, H. Kurita, M. Uota, S. Uchida, M.
Hinomuro and Y. Mera, $J.$ $Vac.$ $Sci.$ $Technol.$ $B$ {\bf 12},
2140 (1994).

\bibitem{12}   F. N. Hooge, $Phys.$ $Lett.$ $A$ {\bf 29}, 139 (1969).



\bibitem{13}   M. Ocio, M. Bouchiat and P. Monod, $J.$ $Magn.$
$Magn.$ $Mater.$ {\bf 54-57}, 11 (1986).

\bibitem{14}   R. H. Koch, W. Reim, A. P. Malozemoff and M. B.
Ketchen, $J.$ $Appl.$ $Phys.$ {\bf 61}, 3678 (1987).

\bibitem{15}    M. B. Weissman and N. E. Israeloff, $J.$
$Appl.$ $Phys.$ {\bf 67}, 4884 (1990).

\bibitem{16}   S. I. Woods, J. R. Kirtley, S. Sun and R. H.
Koch, $Phys.$ $Rev.$ $Lett.$ {\bf 87} 137205 (2001).

\bibitem{17}  V. Podzorov, M. Vehara, M. E. Gershenson, T. Y.
Koo and S. -W. Cheong, $Phys.$ $Rev.$ $B$ {\bf 61} R3784 (2000).

\bibitem{18}   B. Raquet, M. Viret, M. Coster, M. Baibich, M.
Pannetier, M. Blanco-Mantecon, H. Rakoto, A. Maignan, S. Lambert
and C. Fermon, $J.$ $Magn.$ $Magn.$ $Mater.$ {\bf 258-259}, 119
(2003).

\bibitem{19}   T. W. Griswold, A. F. Kip and C. Kittel,
$Phys.$ $Rev.$ {\bf 88} 951 (1952).

\bibitem{20}   G. Feher and A. Kip, $Phys.$ $Rev.$ {\bf 98},
337 (1955).

\bibitem{21}  F. J. Dyson, $Phys.$ $Rev.$ {\bf 98}, 349
(1955).


\bibitem{22}  Y. Manassen, R. J. Hamers, J. E. Demuth and A. J. Castellano
Jr. $Phys.$ $Rev.$ $Lett.$ {\bf 62}, 2531 (1989).

\bibitem{23}  Y. Manassen, E. Ter-Ovanesyan, D. Shachal and S. Richter,
$Phys.$ $Rev.$ $B$ {\bf 48}, 4887 (1993).

\bibitem{24}   Y. Manassen, $J.$ $Magn.$ $Reson.$ {\bf 126}, 133 (1997).

\bibitem{25}   Y. Manassen, I. Mukhopadhyay and N. Ramesh Rao, $Phys.$
$Rev.$ $B$ {\bf 61}, 16223 (2000).

\bibitem{26}   C. Durkan and M. E. Welland, $Appl.$ $Phys.$ $Lett.$ {\bf 80}, 458
(2002).
\bibitem{27} H. C. Manoharan, $Nature$, 416, {\bf 24} (2002).

\bibitem{28}  G. P. Berman, G. W. Brown, M. E. Hawley and V. I.
Tsifernovich, $Phys.$ $Rev.$ $Lett.$ {\bf 87}, 097902-1 (2001).

 \bibitem{29}   D. Mozyrsky, F.
Fedichkin, S. A. Gurvitz and G. P. Berman, $Phys.$ $Rev.$ $B$,
{\bf 66}, 161313 (2002).

\bibitem{30}   A. V. Balatsky, Y. Manassen and R. Salem, $Phil.$ $Mag.$ $B$ {\bf 82},
1291 (2002).

\bibitem{31}   A. V. Balatsky, Y. Manassen and R. Salem, $Phys.$ $Rev.$ $B$ {\bf 66},
195416 (2002).

\bibitem{32}   J. X. Zhu and A. V. Balatsky, $Phys.$ $Rev.$ $Lett.$ {\bf 89}, 286802
(2002).

\bibitem{33}   A. V. Balatsky and I. Martin, cond-mat/01122407

\bibitem{34}   L. S. Levitov and E. I. Rashba, $Phys.$ $Rev.$ ${\bf B 67}$,
115324 (2003)

\bibitem{35} Z. Nussinov, et.al, $Phys.$ $Rev.$ $\bf{ B 68}$,
085402, (2003)

\bibitem{36} L.Bulaevskii, M Hruska  and G. Ortiz,
 $Phys. Rev. {\bf B 68}$, 125415, (2003).

\bibitem{37} A. Shnirman, D. Mozyrsky and I. Martin,
cond-mat/0211618.

\bibitem{38} P.D. Welch, "The use of fast Fourier Transform for
the Estimation of Power Spectra: A Method Based on Time Averaging
over Short Modified Periodgrams", IEEE Trans. Audio. Eletroacoust"
AU-15, 70 (1967).

\bibitem{39} C. Durkan $Contemp.$ $Phys.$ $\bf {45}$,  1, (2004).
\end{thebibliography}
\end{document}